\documentclass[a4,11pt]{article}
\usepackage{epsfig}

\usepackage{amssymb,epsfig,graphicx,epsf,epstopdf}

\usepackage[titletoc,toc,title]{appendix}

\usepackage{amsfonts}
\usepackage{amsmath}
\usepackage{feynmf}

\usepackage{latexsym}
\usepackage{float} 

\usepackage{slashed}
\usepackage{cite}
\usepackage[utf8]{inputenc} 
\usepackage{mathtools}
\usepackage[font=small,skip=-28pt]{caption}
\usepackage{color}
\DeclareFontFamily{OT1}{pzc}{}
\DeclareFontShape{OT1}{pzc}{m}{it}{<-> s * [1.10] pzcmi7t}{}
\DeclareMathAlphabet{\mathpzc}{OT1}{pzc}{m}{it}

\usepackage{commath} 
\usepackage{accents} 
	\newcommand*{\dt}[1]{%
	\accentset{\mbox{\Large\bfseries .}}{#1}} 
\usepackage{relsize} 
\usepackage[makeroom]{cancel} 
\usepackage{xcolor}
\usepackage[normalem]{ulem} 

\usepackage[colorlinks=true, allcolors=black]{hyperref} 

\newcommand{\ben}{\begin{equation}}
	\newcommand{\een}{\end{equation}}
\newcommand{\be}{\begin{equation*}}	
	\newcommand{\ee}{\end{equation*}}

\newcommand{\ba}{\begin{eqnarray}}
\newcommand{\ea}{\end{eqnarray}}
\newcommand{\bal}{\begin{aligned}}
\newcommand{\eal}{\end{aligned}}

\begin{document}

\title{On the propagation of gravitational waves in a $\Lambda$CDM universe}

\author{Jorge Alfaro\\
Pontificia Universidad Cat\'olica de Chile, Av. Vicu\~na Mackenna 4860, Santiago, Chile,\\ 
\\
Dom\`enec Espriu and  Luciano Gabbanelli\\
Departament of Quantum Physics and Astrophysics and\\ 
Institut de Ci\`encies del Cosmos (ICCUB), Universitat de Barcelona,\\
Mart\'\i ~i Franqu\`es 1, 08028 Barcelona, Spain.}

\date{}

\maketitle

\begin{abstract}
We study here how the presence of non-zero matter density and a cosmological constant could affect the observation of 
gravitational waves in Pulsar Timing Arrays. Conventionally, the effect of matter and cosmological constant is included
by considering the redshift in frequency due to the expansion. However, there is an additional effect due to the
change of coordinate systems from the natural ones in the region where waves are produced to the ones used
to measure the pulsar timing residuals. This change is unavoidable as the 
strong gravitational field in a black hole merger distorts clocks and rules. Harmonic waves produced in such a merger 
become anharmonic when detected by a cosmological observer. The effect is small but appears to be observable for the type
of gravitational waves to which PTA are sensitive and for the favoured values of the cosmological parameters. 

\end{abstract}
\vfill
\noindent
November 2017

\noindent
ICCUB-17-022

\newpage

\section{Introduction}

In \cite{EP1,E2} the influence of a non-vanishing cosmological constant on the detection of gravitational waves (GW) in Pulsar
Timing Arrays (PTA) was discussed\footnote{That is, beyond the frequency redshift due to the expansion of the
universe.}. It was found that the value of the time residuals \cite{Espriu} observed in a PTA for a given range of
the angles subtended by the source and the pulsars were dependent on the value of $\Lambda$.
Given that the GW potentially observed in PTA would correspond to the final phase of the fusion of two very massive black holes (BH)
lurking at the center of two colliding galaxies at distances $\sim 1$ Gpc, this result opens the possibility of `local' measurements
of the cosmological constant; namely at subcosmological distances.

To set up the framework for the discussion and avoid misconceptions it is
important to get the correct physical picture from the very beginning. To simplify things, imagine a
large mass $M$ and a much smaller mass $m$ orbiting
around it.  For the time being let us assume that
there is no cosmological constant and no matter density.
In the situation just described, the geometry is very approximately described by the well known Schwarzschild
metric corresponding to a mass $M$. This metric makes use of the radial coordinate $r$ and a time coordinate $t$. This time coordinate $t$
is just the time ticked by a clock located at $r\to \infty$, i.e. an observer at rest at infinity.

Next let us assume that a cosmological constant $\Lambda \neq 0$ is present.
In this case the relevant metric is also well known \cite{Kottler},
	\ben\label{MetricSdS}
	\mathrm{d}s^2=\left(1-\frac{2GM}{r}-\frac{\Lambda}{3}\,r^2\right)\,\mathrm{d}t^2-\left(1-\frac{2GM}{r}-\frac{\Lambda}{3}\,r^2\right)^{-1}\,\mathrm{d}r^2-r^2\,\mathrm{d}\Omega^2\,; \een
where $\mathrm{d}\Omega^2=\mathrm{d}\theta^2+\sin^2\theta\,\mathrm{d}\phi^2$ corresponds to the square of the differential solid angle in spherical coordinates. This metric is unique once the requirements of spherical symmetry and time independence of the metric elements are 
imposed \cite{Birkhoff}.
Here $r$ has the same physical realization as before, but the metric is not anymore Minkowskian when  $r \to \infty$ so one 
must understand carefully what the meaning of that time coordinate is. 

It should be obvious that, in the keplerian problem just described,
if gravitational waves are produced they are periodic in the time $t$ and away from the source they 
will be described by  approximately harmonic functions of the form
\ben\label{gw1}
h^{\scriptscriptstyle (GW)}_{\mu\nu}(w;r,t)=e_{\mu\nu}\, \frac{1}{r}\,\cos\left[w(r-t)\right]\,,\een
with $e_{\mu\nu} $ being the polarization tensor. If at long distances from the source the universe is approximately 
Minkowskian, this wave front will be seen by a remote observer with exactly the same functional form. 
Because of the expansion, our spacetime is not exactly Minkowskian far away from the GW source, but rather described by a
Friedmann--Lemaître--Robertson--Walker (FLRW) metric \cite{FLRW} with coordinates $T$ and $R$
	\ben\label{MetricFRWGeneric}
	\mathrm{d}s^2=\mathrm{d}T^2-a(T)^2\,\left(\mathrm{d}R^2+R^2\mathrm{d}\Omega^2\right).\een
$a(T)$ is the function known as the scale factor. Then a cosmological observer located far away from the BH merger will not see the same functional dependence
in the wave front simply because $T\neq t$ and $R\neq r$. 

In a de Sitter universe the cosmological time and comoving space 
coordinates, the coordinates we use in cosmology and astrophysics, 
are non-trivial functions of the Schwarzschild--de Sitter coordinates $R(t,r)$, $T(t,r)$. These transformations are 
well known and will be reviewed in the next sections.  As a consequence of the different coordinate systems, some 
anharmonicities appear when (\ref{gw1}) is written in terms of our coordinates $T$, $R$. Given that the cosmological 
constant is small, these  anharmonicities are numerically small too, but the results of \cite{EP1} suggest that nevertheless 
they are measurable in a realistic PTA observation.

Conventionally, the expansion of the universe is taken into account by replacing $w$ by its
redshifted counterpart in (\ref{gw1}). If $z$ is small
	\ben w^\prime \simeq  w\,(1-z)\,,\een 
with $z$ being the redshift factor. We will see that this effect on the frequency is indeed reproduced, but there 
is more than this.

The  results obtained in \cite{EP1,E2} were not yet directly applicable to a realistic measurements of PTA time residuals  
because at the very least non-relativistic dust needs to be included to get a realistic description of the cosmology. 
This is in fact the main purpose of the present article.

We have organized our discussion as follows. In the next section we briefly review the notation basic ingredients 
of the $\Lambda$CDM cosmology that will be needed. Next we
discuss 
issues related to the choice of the coordinate system in de Sitter spacetime. This discussion actually contains
the essential ingredient of the present study. 
After this we study the linearization of Einstein's equations and argue why
this is the proper analytical framework to derive the results concerning PTA time residuals. Then we will consider the limiting case
where only dust is present and how the coordinate transformation
between the system of FLRW coordinates and the spherically symmetric coordinates relevant to describe the coalescence of two BH look in 
this case. Finally we combine all the previous results to provide a full answer to the problem. 

In this work we do not enter in any detail in the observational aspects and leave the corresponding analysis for a subsequent paper. 
Here we concentrate on the more fundamental aspects concerning the definition of and relation amongst the different coordinate systems.
We understand that this merits a detailed discussion as many of the results obtained do not appear to be available in the literature.

\section{Conventions and notations in $\Lambda$CDM cosmology}
The Einstein equations, including the cosmological constant and the energy-momentum tensor reads
	\ben\label{EinEq}
	R_{\mu\nu}-\tfrac{1}{2}R\,g_{\mu\nu}-\Lambda g_{\mu\nu}=\kappa\,T_{\mu\nu}\,.\een
The Ricci tensor is constructed by the contraction of the first and third indices of the Riemann curvature 
tensor  $R_{\mu\nu}=R^\lambda\,_{\mu\lambda\nu}$. The Ricci scalar is obtained as usual $R=g^{\mu\nu}R_{\mu\nu}$. The signature convention
for the metric $(+---)$ and $\kappa=8\pi G$. We have included the cosmological constant term $g_{\mu\nu}\Lambda$.

The energy momentum tensor for a perfect fluid in thermodynamic equilibrium takes the form
\ben
T^{\mu\nu}=\left(\rho+P\right)U^\mu U^\nu-P\,g^{\mu\nu}\een
with the chosen convention. In FLRW comoving coordinates $\{T,R,\theta,\phi\}$, it is clear that $U^{T}=U_{T}=1$ whereas $U_{R}=U_\theta=U_\phi=0$.
The normalization condition $g_{\mu\nu}U^\mu U^\nu =1 $ is fulfilled. In order to determine the time evolution of the scale factor, Einstein field equations (\ref{EinEq}) are required together 
with some equation of state that relates the pressure and the density for every single component 
present, $P_i= w_i \rho_i$.

The cosmological constant can be moved, if desired, to the right hand side of Einstein equations and included 
in the form of an energy-momentum tensor 
\ben
T_{\mu\nu}^{\scriptscriptstyle (\Lambda)}=\rho_{ \scriptscriptstyle \Lambda}\,g_{\mu\nu}\,.\een
Here the cosmological constant density $\rho_{ \scriptscriptstyle \Lambda}$ is a constant. In this interpretation, 
the cosmological constant arises from an intrinsic energy density of the vacuum, 
$\rho_{\scriptscriptstyle\Lambda}\equiv \rho_{\scriptscriptstyle vac}$; that is, the new component in the stress-energy tensor can be 
interpreted as an ideal fluid source of energy density which has negative pressure (opposing the pressure of matter if there were) \cite{Peebles}.
Here $w=-1$, hence
\ben\label{CosCteDens&Pressure}
\rho_{\scriptscriptstyle\Lambda}=-p_{\scriptscriptstyle\Lambda}=\frac{\Lambda}{\kappa}\,.\een
By Gauss’s outflux theorem, the cosmological constant is equivalent to a repulsive
gravitational field $\Lambda\,r/3$ away from any center. The current preferred value 
is $\Lambda=\kappa\rho_\Lambda\simeq 10^{-52}\,m^{-2}$ \cite{LambdaValue}.

Using the spatially flat generic metric (\ref{MetricFRWGeneric}) we get from the 1$^{\mathrm{st}}$ Friedmann equation
	\ben\label{ScaleFactorDE}
	\frac{\dt a(T)}{a(T)}\equiv H= \sqrt{\frac{\Lambda}{3}}\,\qquad\Longrightarrow \qquad	a(T)=a_0\,e^{\sqrt{\frac{ \Lambda}{3}}\,\Delta T}\ .\een
Here $H$ is the Hubble constant and the dot stands for derivatives with respect to the cosmological time $T$.
The cosmological constant sets the expansion rate, given by the Hubble parameter $H$. 
$\Delta T=T-T_0$ is the time interval and  $T_0$ is taken such that $a(T_0)=a_0$.

Dust is a special case of a perfect fluid as well. The gravitational field is produced 
entirely by matter characterized by a positive mass density, given by the scalar function $\rho_{\scriptscriptstyle dust}$, but with the
absence of pressure. The stress-energy tensor is 
\begin{equation}\label{EMTensorDust}
	T^{\mu\nu}=\rho_{\scriptscriptstyle dust}\,U^\mu U^\nu\,.\end{equation}
Several independent techniques like cluster mass-to-light ratios \cite{Carlberg-97}, baryon
densities in clusters \cite{Mohr,Grego}, weak lensing of clusters \cite{Mellier,Wilson}, and the existence
of massive clusters at high redshift \cite{ClusterHR} have been used to obtain a handle on $\Omega_{matter}=0.3$. $\Omega_{i}=\rho_{i}/ \rho_{cr}$ with $\rho_{cr}$ being the critical density for which the spatial geometry is Euclidean.
This density is basically dominated by dark matter,
$\rho_{\scriptscriptstyle dust} = \rho_{\scriptscriptstyle DM} +\rho_{\scriptscriptstyle B}$.

As said above, in FLRW comoving coordinates the 4--velocity is 
given by $U^\mu=(1,0,0,0)$; so the stress-energy tensor reduces to $T^{\mu\nu}=\mathrm{diag}(\rho,0,0,0)$.
The Einstein equations derived for this metric structure, together with the equation of state corresponding to $w=0$, give the
solutions\cite{Weinberg} 
\ben\label{ScaleFactorDM}
a(T)=a_0\left(\frac{T}{T_0}\right)^{2/3}\een
and
\ben\label{DensityTRelation}
\rho_{\scriptscriptstyle dust}(T)=\frac{4}{3\,\kappa\,T^2}\,.\een
Therefore, we are able to write the scale factor, and the metric, with a density dependence exclusively
\ben\label{MetricFRWdensity}
\mathrm{d}s^2=\mathrm{d}T^2-a_0^2\left[\frac{\rho_{\scriptscriptstyle d0}}{\rho_{\scriptscriptstyle dust}(T)}\right]^{2/3}\,\left(\mathrm{d}R^2+R^2\mathrm{d}\Omega^2\right).\een
Here, $\rho_{\scriptscriptstyle d0}=4/\bigl(3\,\kappa\,{T_0}^2\bigr)$ is the density measured at a particular cosmological time $T_0$.

Finally, let us consider the combined situation where dust and cosmological constant are both present, and let us derive
the evolution of the corresponding cosmological scale factor in FLRW coordinates. These components are considered as non-interacting fluids; each with such an equation of state. Then follows that the 2$^{\textrm{nd}}$
Friedmann equation holds separately for each such fluid $i$
\ben
\dt\rho_i=-3\,H\,\left(\rho_i+P_i\right)=-3\,H\,\left(1+w_i\right)\rho_i\,;\een
from where we get
	\ben
	\rho_i\propto a^{-3\left(1+w_i\right)}\,.\een
The total density is the addition of the density for each component, using $w_{\scriptscriptstyle\Lambda}=-1$ and $w_{\scriptscriptstyle dust}=0$,
\ben\label{DensityFull}
\rho_{\scriptscriptstyle Full}(T) =\rho_{dust}(T) +\rho_{\scriptscriptstyle\Lambda}=\rho_{d0}\,\left[
\frac{a_0}{a(T)}\right]^3+\rho_{\scriptscriptstyle\Lambda}\,;\een
where $\rho_{d0}$ is the `initial' density of ``dust" and $\rho_{{\scriptscriptstyle\Lambda}}$ is the (constant) density of ``dark energy" when $a(T_0)=a_0$.

To obtain an expression for the scale factor, we have to solve the 1$^{\mathrm{st}}$ Friedmann equation coming from the temporal Einstein field equation combined with (\ref{DensityFull}) 
for the density
\ben\label{Friedmann1stDustCC}
\frac{{\dt a}}{a}=\sqrt{\frac{\kappa\rho_{ \scriptscriptstyle T}}{3}}  =\sqrt{\frac{\kappa\rho_{\scriptscriptstyle d0}}{3}\,\left(\frac{a_0}{a}\right)^3+\frac{\kappa\rho_{\scriptscriptstyle\Lambda}}{3}}\ .\een
Let us pay particular attention on how the `density of dark energy' and the density of dust, are combined at
the level of the equations of motion.  The scale factor is written as
\ben\label{ScaleFactorDMDE}\begin{split}
	a(T)&=a_0\left[\sqrt{1+\frac{\rho_{\scriptscriptstyle d0}}{\rho_{\scriptscriptstyle\Lambda}}}\,\sinh\left(\sqrt{3\,\kappa\rho_{\scriptscriptstyle\Lambda}}\,\frac{\Delta T}{2}\right)+\cosh\left(\sqrt{3\,\kappa\rho_{\scriptscriptstyle\Lambda}}\,\frac{\Delta T}{2}\right)\right]^{2/3}\ .\end{split}\een
In the limit when $\rho_{\scriptscriptstyle d0}\rightarrow0$ we recover the one component scale factor (\ref{ScaleFactorDE});
or if $\rho_{\scriptscriptstyle\Lambda}\rightarrow0$, we get (\ref{ScaleFactorDM}) as it should be. From Eq. (\ref{DensityFull}) together with the
explicit scale factor (\ref{ScaleFactorDMDE}) we find a bijection between $\Delta T$ and $\rho_{\scriptscriptstyle dust}$. Furthermore, the metric
written in terms of the dust density takes exactly the same structure than for a matter dominated universe only; i.e. written in
terms of the density, the metric becomes the same as in (\ref{MetricFRWdensity}), but this time the density for
dust $\rho_{\scriptscriptstyle dust}(T)= \rho_{\scriptscriptstyle Full}(T) -\rho_{\scriptscriptstyle\Lambda}$ involves $\rho_{\scriptscriptstyle\Lambda}$.


\section{The choice of coordinate systems}\label{CosCte}

The de Sitter geometry models a spatially flat universe and neglects matter, so the dynamic of the universe 
is dominated by the cosmological constant $\Lambda$. Obviously there exist a number of useful coordinate choices for this spacetime. 
These consist in picking a convenient time choice and thus defining a family of spacelike surfaces. This is referred 
to as a slicing of the space. Some slices make explicit a cosmologically expanding space with flat curved spatial 
part, whereas some other do not.

There also exists a unique choice of coordinates in which the metric does not depend on time at all; it is the one
described in Eq. \eqref{MetricSdS} if the BH contribution is omitted
\ben\label{MetricdS}
\mathrm{d}s^2=\left(1-\frac{\Lambda}{3}\,r^2\right)\,\mathrm{d}t^2
-\left(1-\frac{\Lambda}{3}\,r^2\right)^{-1}\,\mathrm{d}r^2-r^2\,\mathrm{d}\Omega^2. 
\een
The mere existence of such a choice 
is enough to tell us that there is no fundamental sense
in which this is an expanding cosmological spacetime. The notion of expansion is a concept that is linked to a coordinate choice.
Yet, the choice of coordinates is not totally arbitrary. Observers' clocks tick at a given rate, and they can measure the local
space geometry at a fixed time. 

Cosmological observers detect that the universe as we see it is spatially flat and that
unbound free-falling objects separate from each other at a rate that is governed by 
the cosmological scale factor $a(T)$ present in Eq. \eqref{MetricFRWGeneric}.
Then the physical coordinate system for discussing cosmology is given by the FLRW metric, as
this coordinate system encodes the observed homogeneity and isotropy of the matter sources summarized in the cosmological principle, 
valid when the universe is viewed on a large enough scale \cite{Book}.  
Of course, space redefinitions of the metric are innocuous. It is of no consequence to choose to describe the world around 
us using Cartesian or polar coordinates. As long as coordinate transformations do not involve time in any essential way, 
any coordinate choice will be equally good to describe a universe conforming to the cosmological principle.

Time and space will however look very different to an observer in the vicinity of a BH or, for that matter, in the vicinity of any
strong gravitational source that is spherically symmetric centered at one point that
we conventionally denote by $r=0$. The spacetime there is isotropic, but not homogeneous. If the BH is static, we know that the solution
is unique and it is given by the Schwarzschild metric. If in addition, there is a cosmological constant, the corresponding metric 
will be the one given in Eq. (\ref{MetricSdS}). Note that this metric exhibits two horizons: one at $r=r_S=2GM$ and another
one at $r_\Lambda=\sqrt{3/\Lambda}$. 
These coordinates are named Schwarzschild--de Sitter (SdS) and the corresponding metric is the one in Eq. (\ref{MetricSdS}). 
We will label these coordinates with lowercase letters, $\{t,r,\theta,\phi\}$.
Then, the SdS metric approximates a Schwarzschild space for `small' $r$; and for ‘large’ $r$ the
space approximates a de Sitter space. We are focusing right now on effects on gravitational wave propagation, hence we are in
a ‘large’ $r$ regime.

Away from the BH, i.e. for $r\gg r_s$  the $1/r$ dependence can be safely neglected from our considerations and only 
the $\Lambda $ dependence needs to be taken into account; i.e. under this regime the metric is given by \eqref{MetricdS}. In this case the transformation relating the SdS coordinates 
to the FLRW ones are 
well known (see e.g. \cite{Espriu})
\ben\label{ChVarCCExact}
t(T,R)=T-\sqrt{\frac{3}{\Lambda}}\,\log\sqrt{1-\frac{\Lambda}{3}\,a^2(T)\,R^2}\hspace{8pt};
\hspace{30pt} r(T,R)=a(T)\,R=a_0\,e^{\sqrt{\frac{\Lambda}{3}}\,\Delta T}\,R \,;\een
the $\theta$ and $\phi$ coordinates do not transform. Note that even though for $r\gg r_S$ the presence of the BH 
is negligible, its presence sets up a global coordinate system, and $r$ and $t$ have a well defined physical meaning.

A linearized Schwarzschild-de Sitter metric is obtained after expanding (\ref{MetricdS}) for a small cosmological constant $\Lambda\ll1$.
For $r\to \infty$
	\ben\label{MetricSdSCCAprox}
	\mathrm{d}s^2=\left(1-\frac{\Lambda}{3}\,r^2\right)\,\mathrm{d}t^2-\left(1+\frac{\Lambda}{3}\,r^2\right)\,\mathrm{d}r^2-r^2\,\mathrm{d}\Omega^2.\een
This metric satisfies a linearized version of Einstein field equations (\ref{EinEq}).
On the contrary, the FLRW metric does not fulfill any linearized version of Einstein
equations\footnote{This is easy to understand by realizing that in the FLRW metric $\Lambda$ appears in a non-analytic form.};
in fact, no metric that depends only on time can be solution of the linearized Einstein equations whatever the gauge choice selected. 

The previous coordinate transformations can be expanded for small values of $\Lambda T^2$ (this will always be the relevant situation for
the ensuing discussions)
	\ben\label{ChVarCCLinear}\left\{\begin{split}
	\hspace{7pt} t(T,R)&\approx
	T+a_0\left(\frac{R^2}{2}\sqrt{\frac{\Lambda}{3}}+R^2\,\Delta T\,\frac{\Lambda}{3}\right)+\dots
	\\[2ex]
	r(T,R)
	&\approx
	a_0\left[1+\Delta T\,\sqrt{\frac{\Lambda}{3}}+\Delta T^2\,\frac{\Lambda}{6}\right]R+\dots\end{split}\right.\een
The change is of course still non--analytical in $\Lambda$. Applying this approximate transformation to the SdS metric, one
ends up with an approximate version of the FLRW metric
\ben
	\mathrm{d}s^2\approx\mathrm{d}T^2-a_0^2\left(1+2\,\sqrt{\frac{\Lambda}{3}}\,\Delta T+2\,\frac{\Lambda}{3}\,\Delta T^2\right)\left(\mathrm{d}R^2+R^2\,\mathrm{d}\Omega^2\right)\een
which is the expansion of (\ref{MetricFRWGeneric}) for small values of the cosmological constant, $\Lambda r^2<<1$. In practice, the leading $\sqrt\Lambda$ term
will be the relevant one for our purposes.


\section{Why linearized gravity}
The linearized theory of gravity is nothing else than perturbation theory around Minkowski spacetime. 
This is, starting with a small deviation  $\abs{h_{\mu\nu}}\ll1$ from a flat metric tensor,
\ben\label{MetricetricPertMink}
	g_{\mu\nu}=\eta_{\mu\nu}+h_{\mu\nu}\,.\een
The usual procedure for obtaining the linearized Einstein tensor can be found in any 
textbook on general relativity (see e.g. \cite{Cheng}). However, a gauge choice is mandatory to obtain a solution to 
these equations in order to avoid redundancy under coordinate transformations $x^\mu\rightarrow x^\mu+\xi^{\mu}(x)$. 
Although the following discussion is valid in any gauge, it is simplest in the familiar
Lorenz gauge
\ben\label{LorenzGaugeCond1}
\partial_\mu h^\mu{}_\nu - \frac{1}{2}\partial_\nu h=0 \,,\een
or defining
\ben\label{LorenzGaugeCond2}
\tilde h_{\mu\nu}=  h_{\mu\nu} - \frac{1}{2}\eta_{\mu\nu} h\quad \Rightarrow \quad
\partial_\mu \tilde h^\mu{}_\nu =0\,.\een
In this gauge the linearized Einstein equations, including the cosmological constant and the energy momentum tensor read
\ben\label{FullLin}
	\Box \tilde h_{\mu\nu}=-2\,\Lambda\eta_{\mu\nu} -2  \kappa\,T_{\mu\nu}\,.\een
As discussed in \cite{Espriu}, up to first order in $\Lambda$ the perturbation $h_{\mu\nu}$ 
in Eq. (\ref{MetricetricPertMink}) can be decomposed into a gravitational wave perturbation $h_{\mu\nu}^{\scriptscriptstyle (\mathrm{GW})}$ 
and a background modification due to the cosmological constant $h_{\mu\nu}^{\scriptscriptstyle (\Lambda)}$; being both `small' 
in magnitude. It is straightforward to propose a new contribution due to dust $h_{\mu\nu}^{\scriptscriptstyle (\mathrm{dust})}$. 
This component would contribute as an independent new background that will not interact with the other one, the one 
corresponding to $\Lambda$, and that will be also `small' enough so that the linearized approximation to be meaningful. 
Therefore, the total metric would be written as
	\ben
	g_{\mu\nu}=\eta_{\mu\nu}+h_{\mu\nu}^{\scriptscriptstyle (\mathrm{GW})}+h_{\mu\nu}^{\scriptscriptstyle (\Lambda)}+h_{\mu\nu}^{\scriptscriptstyle (\mathrm{dust})}\,.\een
The wave equation will have the same structure but the perturbation would have the three contributions 
mentioned before. The first two contributions has been already discussed in \cite{Bernabeu}. 

The two independent background would satisfy each corresponding equation of motion
\ben\label{LinLam}
	\Box \tilde h_{\mu\nu}^{\scriptscriptstyle (\Lambda)}=-2\,\Lambda\eta_{\mu\nu}\een
for the cosmological constant and 
\ben
	\Box \tilde h_{\mu\nu}^{\scriptscriptstyle (\mathrm{dust})}=-2\kappa T_{\mu\nu}\een
for a dark matter background uniformly distributed along the spacetime. Finally the perturbation due to
the gravitational wave itself satisfies the usual homogeneous wave equation
\ben
	\Box \tilde h_{\mu\nu}^{\scriptscriptstyle (GW)}=0\,.\een
Thus the discussion is simple in linearized gravity. The contribution to the different terms simply add to construct the
total metric.

The solution of Eq \eqref{LinLam} as explained in \cite{Espriu} should correspond to the 
linearization (i.e. expanding in $\Lambda$ at first order)
of Eq. \eqref{MetricdS}; i.e. the metric \eqref{MetricSdSCCAprox}. In fact this statement is not totally correct because in (\ref{FullLin}) the Lorenz gauge condition
is assumed, while the SdS metric does not fulfill this gauge condition, as it can be easily checked.
In fact, as explained in \cite{Espriu} a trivial
coordinate transformation turns the SdS metric into one that complies with the Lorenz gauge. This coordinate
transformation is (a) time-independent and (b) is of order $\Lambda$. The discussion on GW and the separation
in background and waves is however simplest in the Lorenz gauge and it is the one just presented. In fact, the relevant
effects to be discussed later are all of order $\sqrt{\Lambda}$ and order $\Lambda$ corrections are safe to neglect.


\section{Relevance for Pulsar Timing Arrays}\label{PTA}
After transformation to FLRW coordinates, and keeping the 
changes of order $\sqrt\Lambda$ only, Eq. (\ref{gw1}) reads
\ben\label{gwfrw}
h{}^{\scriptscriptstyle (FLRW)}_{\mu\nu}\equiv h'{}^{\scriptscriptstyle (GW)}_{\mu\nu}\simeq 
\frac{e^\prime_{\mu\nu}}{R}\left( 1+\sqrt{\frac{\Lambda}{3}}\,T\right)
\cos\left[w\,(T-R)+w\,\sqrt{\frac{\Lambda}{3}}\left(\frac{R^2}{2}-TR\right)\right]\,.\een
$e^\prime_{\mu\nu}$ is the transformed polarization tensor that will not be very relevant for the following discussion, although its precise form is essential for precise comparison with observations.

This last expression can also be written as
\ben\label{gwfrw3}
h{}^{\scriptscriptstyle (FLRW)}_{\mu\nu}=\frac{e^\prime_{\mu\nu}}{R}\left( 1+\sqrt{\frac{\Lambda}{3}}T\right)
\cos\left[w_{\scriptstyle eff}T-k_{\scriptstyle eff}R \right]\,,\een
with $w_{\scriptstyle eff}= w\bigl(1-R\sqrt{\Lambda/3}\bigr)$ and $k_{\scriptstyle eff}= w\bigl(1- R/2\sqrt{\Lambda/3}\bigr)$.
Notice that $w_{\scriptstyle eff}$ agrees with the usual frequency redshift, so this well known result is well reproduced. However
the wave number is different. This discrepancy is at the root of the observability of $\Lambda$.

As emphasized before, the corrections of order $\Lambda$ and beyond are really irrelevant when the actual presumed value for the cosmological constant is considered; only $\sqrt\Lambda$ terms really matter. In the Lorenz gauge the 
only spatial components of the metric that are different from zero are 
the $X,Y$ entries of $e^\prime_{\mu\nu}$. Although some temporal components are 
also non-zero in these coordinates, they are several orders of magnitude 
smaller than the spatial ones and therefore will not be relevant for the 
present study.

The phase velocity of propagation of the GW in such coordinates 
is  $v_{p}\sim 1-\sqrt{\Lambda/3}\,T + \mathcal{O}(\Lambda)$ \cite{Espriu}. 
On the other hand, with respect to the ruler distance traveled (computed with $g_{ij}$) the velocity is still $1$.

Consider the situation shown in Figure 1 describing the relative situation of a GW source 
(possibly a very massive black hole binary), the Earth and a nearby pulsar.
\begin{figure}[htb]
	\begin{center}
	\includegraphics[scale=0.5]{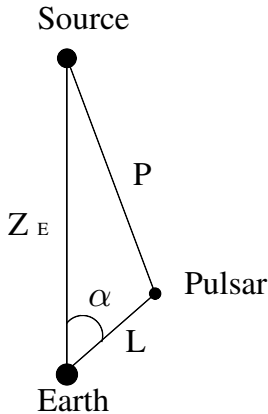}
	\vspace{40pt}\captionsetup{width=0.9\textwidth}
	\caption{Relative coordinates of the GW source ($R=0$), the Earth (at $R=Z_E$ with respect to the GW source)
	and the pulsar (located at coordinates $\vec P= (P_X, P_Y, P_Z)$ referred to the source). In FLRW coordinates centered on Earth, the angles $\alpha$ and $\beta$ are the polar and azimuthal angles of the pulsar with respect to the axis
	defined by the Earth-source direction.}\label{coor} \end{center}\end{figure}
The timing residual induced by (\ref{gwfrw}) will be given by the integral
\ben\label{tim}
H(T_E,L,\alpha,\beta,Z_E,w,\varepsilon,\Lambda)= -\frac{L}{2c} \hat n^i \hat n^j \int_{-1}^0\ dx\ h_{ij}^{\scriptscriptstyle(FLRW)}\left[T_E +\frac{L}{c}x,
\vec P + L(1 + x)\hat n\right]\een
along the null geodesic from the pulsar to the earth $\vec R(x)= \vec{P}+L(1+x)\hat{n}$. The unit vector $\hat n$ is given by $(-\sin\alpha \cos\beta,-\sin\alpha \sin\beta, \cos\alpha)$, $Z_{E}$ is the distance from 
Earth to the GW source, $L$ the distance to the pulsar, $T_{E}$ the time of arrival of the wave to the local 
system and $\varepsilon \sim | e^\prime_{ij}|$, $i,j=X,Y$. In deriving the previous timing residual we have neglected
the peculiar motion of the Earth. The speed of light has been restored in this formula. We have 
assumed that from the pulsar to the Earth the electromagnetic signal follows the trajectory given by the line of sight
and is the distortion created by the intervening GW what modifies the timing residual.

\begin{figure}[!h]
	\begin{center}
	\includegraphics[scale=0.47]{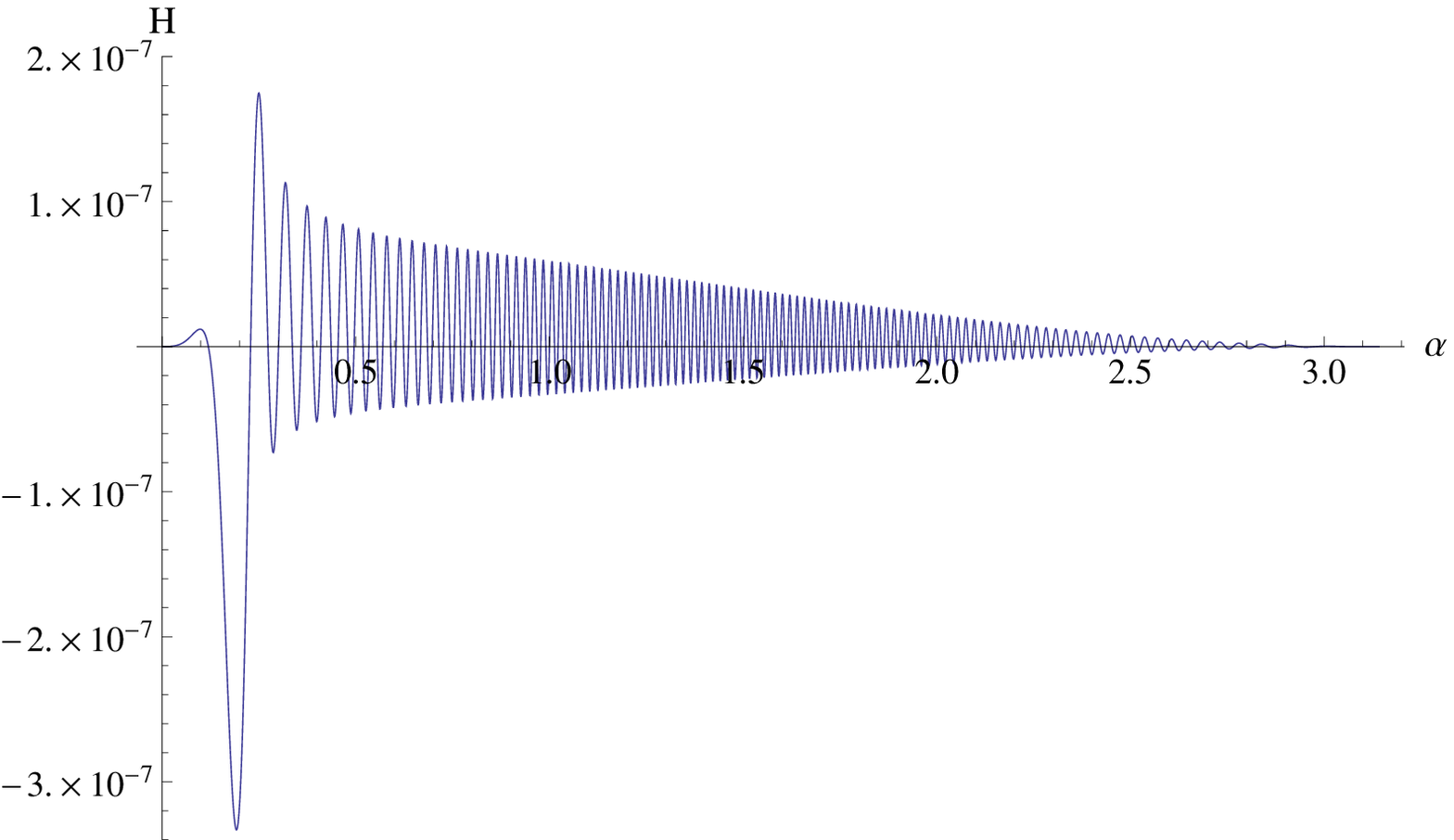}
	\includegraphics[scale=0.47]{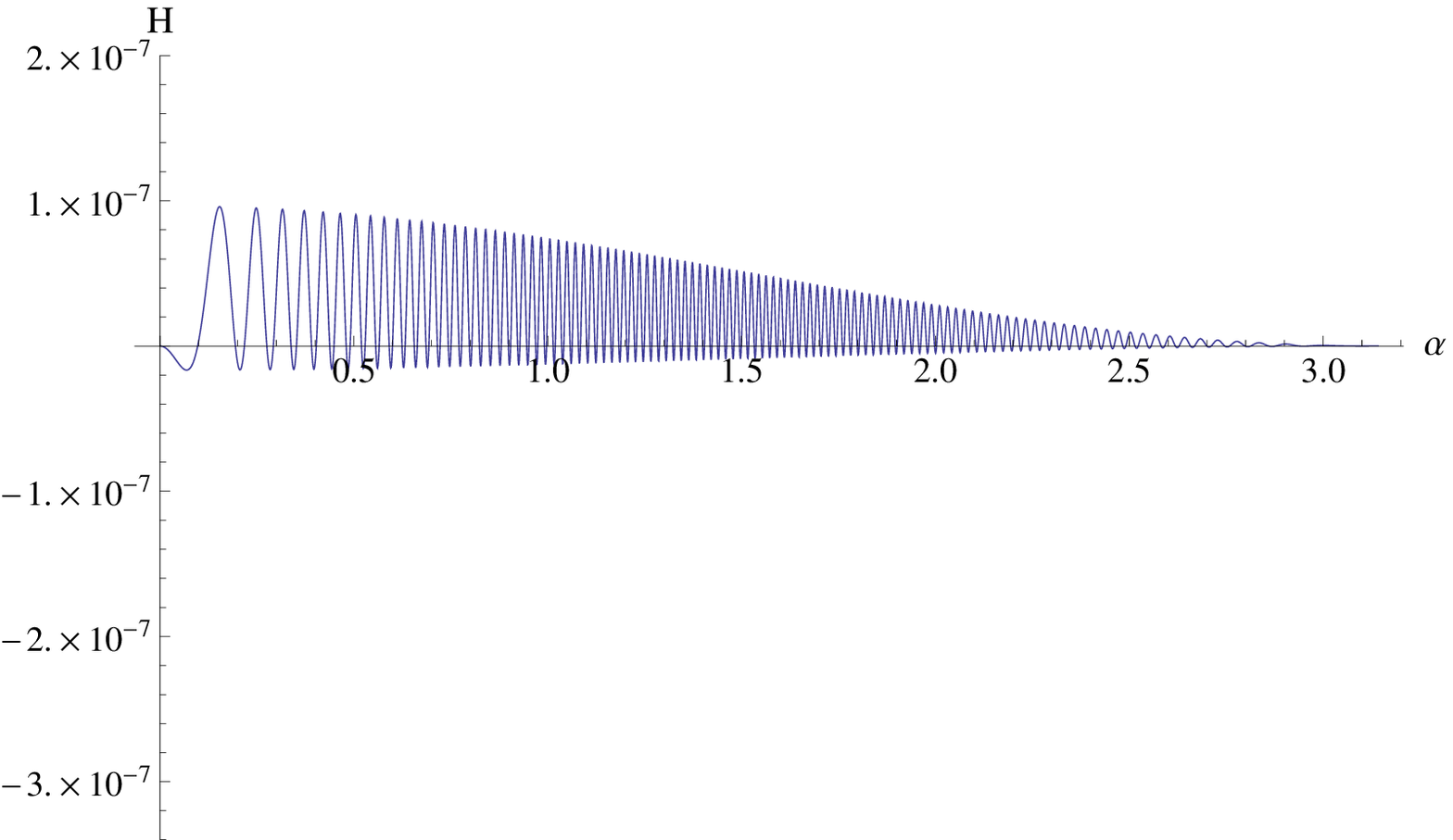}
	\vspace{40pt}\captionsetup{width=0.9\textwidth}
	\caption{On the left the raw timing residual for $\Lambda=10^{-35}s^{-2}$ as a function
	of the angle $\alpha$ subtended by the source and the measured pulsar as seen from the observer. The figures are symmetrical for $\pi\leq\alpha\leq 2\pi$.
	On the right the same timing 
	residual for $\Lambda=0$. In both cases we take $\varepsilon=1.2\times 10^{9}m$, $L=10^{19}m$ and $T_{E}=\frac{Z_{E}}{c}s$ for $Z_{E}=3\times 10^{24}m$; with these values $|h|\sim \frac{\varepsilon}{R}\sim 10^{-15}$ which is
	within the expected accuracy of PTA \cite{jenet}. Similar results are obtained for other close values of $T_{E}$. From reference \cite{EP1}.}\label{time2}\end{center}\end{figure}

The real question is whether the observationally preferred exceedingly small value of the 
cosmological constant affects the timing residuals from a pulsar at all. 
This question was answered in the affirmative in \cite{EP1}. We take reasonable 
values of the parameters both for the GW and one pulsar location (quoted in the caption) and plot a snapshot of the 
resulting timing residuals as a 
function of the angle $\alpha$ for the time of arrival of the wave to the local system, $T_{E}$. In Figure \ref{time2}
we compare the cases with and without cosmological constant.

The figure speaks by itself and it strongly suggests that the angular dependency of the timing 
residual is somehow influenced by the value of the
cosmological constant, in spite of its small value. Another feature that catches the eye immediately is an 
enhancement of the signal for a specific small angle $\alpha$, corresponding generally to a source of low
galactic latitude or a pulsar nearly aligned with the source (but not quite as otherwise $e^\prime_{ij}\hat n^{i}\hat n^{j}=0$, although the total 
timing residual is non-vanishing due to the $\mathcal{O}(\Lambda)$ time components for waves) . 

This enhancement is relatively easy to understand after a careful analysis of the integral in Eq. (\ref{tim}), as
was explained in \cite{EP1}. However for us it is only necessary to know that the effect is visible and therefore
the need to consider carefully the different coordinate systems where GW are produced and measured.

In the rest of the paper we want to analyze the possible modifications that the presence of dust can bring about.


\section{Non-relativistic matter}\label{NoRelMatt}

Now we would like to repeat the discussion done in the previous sections for a purely de Sitter universe in the case
where there is no cosmological constant, just dust. Namely
we ask ourselves what would be the equivalent of Eq. (\ref{ChVarCCExact}) in such a case.

With the aim of deriving a full picture on how dust influences the observation of gravitational waves, we need to find
the coordinate system where dust is described in `static' spherically symmetric coordinates with an origin coincident with the object
emitting gravitational waves.
To make things clear, dust only plays a role in the definition of such a coordinate system
and its relation with the observer's reference frame, namely FLRW coordinates.

The first step is to find a set of coordinates, spherically symmetric, where the angular element that in FLRW reads
$a(T)^2 R^2\mathrm{d}\Omega^2$ should become simply $r^2 \mathrm{d}\Omega^2$. 
The coordinate transformation will be given by
\ben\label{ChVarDMUnknown}\left\{\,\begin{split}
	\hspace{7pt}T(t,r)&=\text{unknown function}\hspace{40pt} \\
	R(t,r)&=\frac{r}{a(T)}\end{split}\right.\een

Next we take into account the transformation properties of a rank 2 tensor
\ben\label{MetricTransformation}
	g^\prime_{\mu'\nu'}=\frac{\partial X^\mu}{\partial{x^\mu}'}\,\frac{\partial X^\nu}{\partial{x^\nu}'}\,g_{\mu\nu}\,.\een
Spherical symmetry is a requirement for both metrics, then the transformation for the two angular variables is characterized by the identity.
This means that in the angular diagonal component of the metric the Jacobian corresponds to the identity and the transformation for
these component do not bring any new information. Analogous is the case for the angular and off-diagonal elements of the metric.
The transformation is trivial and in both pictures the components of the metric are null. However, there is one off-diagonal element
that is not null \textit{per se}; the $g_{tr}=g_{rt}$ element. If a diagonal metric is desired we must impose the vanishing of this element.
This translates into the following differential equation for the cosmological time $T$
	\ben\label{DiagonalMetricCondition}
	\partial_r T=-\frac{6\,r\,T}{9T^2-4\,r^2} \qquad\iff\qquad\partial_r\rho_{\scriptscriptstyle dust}= \frac{\kappa\,\rho_{\scriptscriptstyle dust}^2\,r}{1-\frac{\kappa\,\rho_{\scriptscriptstyle dust}}{3}\,r^2}\,.\een
Of course as we already know the structure for the bijection between the cosmological time and the content of the universe,
i.e. $\abs{T}\propto1/\sqrt{\rho_{\scriptscriptstyle dust}}$ coming from Eq. (\ref{DensityTRelation}). This allows to write
the diagonal metric condition (\ref{DiagonalMetricCondition}) in terms of the non-relativistic matter density only.
We have two remaining components of the metric to be transformed: the $TT$ and $RR$--components. If we use only the
requirement for the radial transformation (\ref{ChVarDMUnknown}) and impose a diagonal structure for the metric (\ref{DiagonalMetricCondition}), the metric reads
	\ben\label{MetricSSSDMUnknown}
	\mathrm{d}s^2=\frac{\left(\partial_t\rho_{\scriptscriptstyle dust}\right)^2}{3\,\kappa\,{ \rho_{\scriptscriptstyle dust}}^3} \left[1-\frac{\kappa\,\rho_{\scriptscriptstyle dust}}{3}\,r^2\right]\,\mathrm{d}t^2- \frac{1}{1-\frac{\kappa\,\rho_{\scriptscriptstyle dust}}{3}\,r^2}\,\mathrm{d}r^2 -r^2\,\mathrm{d}\Omega^2\,.\een
Note that $\rho_{\scriptscriptstyle dust}$ is a (scalar) known function of $T$ and therefore it is expected to depend on the
new temporal variable $t$; but this dependence is unknown a priori (because $t$ is unknown so far). In any case we see
that the new metric necessarily contains a $t$ dependence via $\rho_{\scriptscriptstyle dust}(t)$.

Before continuing, let us reduce the number of dimensionful parameters of the theory. The limit of weak coupling ($\kappa\rightarrow0$)
and vanishing matter density ($\rho_{\scriptscriptstyle dust}\rightarrow0$) are essentially the same because both parameters appear
together in all the equations --except in the energy momentum conservation which is trivially satisfied. Hence we
define: $\widetilde{\rho}_{\scriptscriptstyle dust}=\kappa\rho_{\scriptscriptstyle dust}$. Then the solution of (\ref{DiagonalMetricCondition}) is
	\ben\label{RelationSolvesODE}
	\frac{6+r^2\,\widetilde{\rho}_{\scriptscriptstyle dust}}{{\widetilde\rho_{\scriptscriptstyle dust}}^{1/3}}=C(t)\,;\een
where $C(t)$ is a constant with respect to $r$ meaning that it can only depends on $t$. Dimensional analysis do the rest of the work.
In natural units, $[\widetilde\rho_{\scriptscriptstyle dust}]=L^{-2}$, then the units for the `constant' should
be $[C(t)]=[\widetilde\rho_{\scriptscriptstyle dust}]^{-1/3}=L^{2/3}$. In this picture, $\kappa$ does not belong to the theory, hence there
is only one dimensionful parameter to give units to $C(t)$; namely $t$ because the constant is radial independent. As $\left[\,t\,\right]=L$,
then $C(t)=A\,t^{2/3}$ with $A$ a positive dimensionless parameter to be determined.

It is expected for later times the dust density to be homogeneously diluted. This requirement automatically fixes the remaining
free parameter of the theory, $A$. This is translated into the metric (\ref{MetricSSSDMUnknown}) to be Minkowskian--like when $t\rightarrow\infty$.
This is automatically fulfilled except for the factor in front of the brackets in the $g_{tt}$ element. If a Minkowskian limit 
is expected, the following enforcement of the prefactor is needed
$(\partial_t\rho_{\scriptscriptstyle dust})^2/(3\kappa{\rho_{\scriptscriptstyle dust}}^3)\rightarrow1$. We have a relation for obtaining the temporal variation of the dust density, namely (\ref{RelationSolvesODE}) with the corresponding $C(t)$ function. Therefore, reintroducing $\kappa$ and deriving both sides of the equality, one arrives to
	\ben
	\partial_t(\kappa\rho_{\scriptscriptstyle dust})=-\frac{(\kappa\rho_{\scriptscriptstyle dust})^{4/3}}{1-\frac{\kappa\rho_{\scriptscriptstyle dust}}{3}\,r^2} \,\frac{A}{3\,t^{1/3}}\,.\een
Squaring the last result, dividing by $3\,(\kappa\rho_{\scriptscriptstyle dust})^3$, and finally using Eq. (\ref{RelationSolvesODE})
again for replacing $1/t^{2/3}$, the prefactor for the $g_{tt}$ component in the limit of $\rho_{\scriptscriptstyle dust}\rightarrow0$ becomes
	\ben
	\frac{A^2}{3^3\,(\kappa\rho)^{1/3}\,t^{2/3}}\to \frac{A^3}{3^3\,6}\, .
	\een
After all, imposing the flat asymptotic limit to this factor, the constant is fixed to $A=3\sqrt[3]{6}$. Once $A$ is known, 
$\partial_t\rho_{\scriptscriptstyle dust}$ can be expressed as a function of $\rho_{\scriptscriptstyle dust}$ and $r$ uniquely. Following the same steps as before; replacing $1/t^{2/3}$ from the prefactor by $A/C(t)$ from Eq. \eqref{RelationSolvesODE}, we get
simply
	\be
	\frac{\left(\partial_t\rho_{\scriptscriptstyle dust}\right)^2}{3\,\kappa\,{ \rho_{\scriptscriptstyle dust}}^3}= \frac{1}{\left(1+\frac{\kappa\,\rho_{\scriptscriptstyle dust}}{6}\,r^2\right)\left(1-\frac{\kappa\rho_{\scriptscriptstyle dust}}{3}\,r^2\right)^2}\ee
where the asymptotic limit is shown explicitly.
Then the sought for metric turns out ot be
	\ben\label{MetricSSSDM}
	\mathrm{d}s^2=\frac{1}{\left(1+\frac{\kappa\,\rho_{\scriptscriptstyle dust}}{6}\,r^2\right)\left(1-\frac{\kappa\,\rho_{\scriptscriptstyle dust}}{3}\,r^2\right)}\,\mathrm{d}t^2 -\frac{1}{1-\frac{\kappa\,\rho_{\scriptscriptstyle dust}}{3}\,r^2}\,\mathrm{d}r^2-r^2\,\mathrm{d}\Omega^2\,.\een
This metric has a good Minkowskian limit for $\rho_{\scriptscriptstyle dust}\rightarrow0$. It has similar characteristics to the ones of the
SdS metric and it is clearly the relevant metric to describe the Keplerian problem we alluded to in the
introduction in the presence of dust, but without cosmological constant, when properly extended with the BH gravitational potential (see below).

Further discussions concerning dust in these coordinates are included in the appendix. With the full structure for $C(t)$, we present a detailed discussion on the physical solutions to Eq. \eqref{RelationSolvesODE}. In appendix \ref{DustDensity} we show that the mere existence of a solution for \eqref{RelationSolvesODE} entails the presence of a horizon in such coordinates. 
In appendix \ref{DustEnergyMomentumTensor}, we include the full computation of the \textit{non-diagonal} stress-energy tensor also in these coordinates .

The corresponding change of variables to and from FLRW is given by
\ben\label{ChVarDMExact}\left\{\begin{split}	
	&\quad t=
	\frac{\left[6 +(\kappa\rho_{\scriptscriptstyle d0})^{2/3}\,(\kappa\rho_{\scriptscriptstyle dust})^{1/3}\,R^2 \right]^{3/2}}{9\,\sqrt{2\,\kappa\rho_{\scriptscriptstyle dust}}}
	\\
	&\quad r=\sqrt[3]{\frac{\rho_{\scriptscriptstyle d0}}{\rho_{\scriptscriptstyle dust}}}\,R \end{split}\right.\een
with $\rho_{\scriptscriptstyle dust}$ given by Eq. (\ref{DensityTRelation}).
The metric (\ref{MetricSSSDM}) can be linearized for $\rho_{\scriptscriptstyle dust}\to 0$ to
	\ben\label{MetricSSSDMLinear}
	\mathrm{d}s^2=\left(1+\frac{\kappa\,\rho_{\scriptscriptstyle dust}}{6}\,r^2\right)\,\mathrm{d}t^2-\left(1+\frac{\kappa\,\rho_{\scriptscriptstyle dust}}{3}\,r^2\right)\,\mathrm{d}r^2-r^2\,\mathrm{d}\Omega^2\,.\een
Within this linearized approximation it is of course trivial to include the BH gravitational field. This would give rise to
	\ben\label{MetricSSSDMLinearplusBH}
	\mathrm{d}s^2=\left(1-\frac{r_S}{r}+\frac{\kappa\,\rho_{\scriptscriptstyle dust}}{6}\,r^2\right)\,\mathrm{d}t^2-\left(1+\frac{r_S}{r}+\frac{\kappa\,\rho_{\scriptscriptstyle dust}}{3}\,r^2\right)\,\mathrm{d}r^2-r^2\,\mathrm{d}\Omega^2.\een
This expression is valid in the region $r\gg r_S$ and $\kappa\rho_{\scriptscriptstyle dust}\,r^2 \ll 1$. Note that it is not
time-independent, unlike its SdS counterpart, but the time dependence enters only via the matter density.

In Eq. (\ref{ChVarDMExact}) the coordinate transformation is expressed in terms of $\rho_{\scriptscriptstyle dust}$ but the dependence  of $\rho_{\scriptscriptstyle dust}$
on $T$ is known (see Eq. (\ref{DensityTRelation})). We can therefore expand in powers of $\Delta T=T-T_0$, where $T_0$ is the time
where the density of dust is $\rho_{\scriptscriptstyle d0}$. 
Linearization of the coordinate transformation leads to
\ben\label{ChangeVariablesLinearDustTrue}\left\{\begin{split}
	&\quad t=T+\frac{1}{2}\,\sqrt{\frac{\kappa\rho_{\scriptscriptstyle d0}}{3}}\,R^2
	+\frac{\kappa\rho_{\scriptscriptstyle d0}}{12}\,\Delta T\,R^2+\dots
	\\
	&\quad r=\left(1+\sqrt{\frac{\kappa \rho_{\scriptscriptstyle d0}}{3}}\,\Delta T-\frac{1}{4}\frac{\kappa\rho_{\scriptscriptstyle d0}}{3}\,\Delta T^2\right)\,R+\dots\end{split} \right.\een
where the scale factor $a_0$ is taken equal to 1 from now on.
Note that the initial condition $T=T_0$ for the cosmological time does not correspond to a unique value 
for $t$ as the relation does depend on $R$. 

Now we have all the ingredients to combine both components, dust and cosmological constant, as described in 
sections \ref{CosCte} and \ref{NoRelMatt}.


\section{Non-relativistic matter and cosmological constant}\label{DMandCC}
Now we are in a position to put together the two ingredients, namely dust and cosmological constant. In 
linearized gravity this is exceedingly simple as previously discussed. The metric will be 
\ben\label{MetricSSSDMCCLinear}
	\mathrm{d}s^2=\left(1+\frac{\kappa\rho_{\scriptscriptstyle dust}}{6}\,r^2-\frac{\Lambda}{3}\,r^2\right)\mathrm{d}t^2-\left(1+\frac{\kappa\rho_{\scriptscriptstyle dust}}{3}\,r^2+\frac{\Lambda}{3}\,r^2\right)\mathrm{d}r^2-r^2\,\mathrm{d}\Omega\,,\een
where the BH contribution $\sim r_S/r$ has been omitted because is not relevant at large distances.
The additive picture permits to recover the individual linearized theory for each component (taking the corresponding limit of the other going to zero). One would expect the same additive behaviour in the perturbations for the change of variables, but if one analyze how the full scale factor (\ref{ScaleFactorDMDE})
expands in Taylor series,
\ben\label{ScaleFactorDMCCExpanded}\begin{split}
	a(T)&\simeq1+\sqrt{\frac{\Lambda+\kappa\rho_{\scriptscriptstyle d0}}{3}}\,\Delta T+\dots\,,\end{split} \een
it is clear that the additivity takes place inside the square root. Indeed it is easy to see that the following 
coordinate transformation does the job of moving from FLRW coordinates to the ones in 
Eq. (\ref{MetricSSSDMCCLinear})
\ben\label{ChVarDMCCLinear}\left\{\begin{split}
	\quad&t\sim T+\frac{1}{2}\,\sqrt{\frac{\Lambda+\kappa\rho_{\scriptscriptstyle d0}}{3}}\,R^2+\left( \frac{\Lambda}{3}+\frac{\kappa\rho_{\scriptscriptstyle d0}}{12}\right)\,\Delta T\,R^2+\dots
	\\
	\quad&r=a(T)R\sim\left[1+\sqrt{\frac{\Lambda+\kappa\rho_{\scriptscriptstyle d0}}{3}}\,\Delta T+\dots \right]\,R \end{split}\right.\een
Of course in both limits, $\rho_{ \scriptscriptstyle d0}\rightarrow0$ and $\Lambda\rightarrow0$, we recover 
each 1-component universe coordinate transformation. 

As abundantly mentioned before we will only need the $\sqrt{(\Lambda + \kappa\rho_{ \scriptscriptstyle d0})/3}$ correction
because higher powers of this square root can be neglected.
The result is rather compact and we immediately know the form that a GW adopts in FLRW coordinates
\ben\label{gwfrw2}
h'{}^{\scriptscriptstyle (GW)}_{\mu\nu}=\frac{e^\prime_{\mu\nu}}{R}\left( 1+\sqrt{\frac{\Lambda}{3}}\,T\right)
\cos\left[{w(T-R)+w\sqrt{\frac{\Lambda+\kappa \rho_{\scriptscriptstyle d0}}{3}}\left(\frac{R^2}{2}-TR\right)}\right]\,.\een
This implies
\ben\label{effconstants}
w_{\scriptstyle eff}= w \left(1- R \sqrt{\frac{\Lambda+\kappa \rho_{\scriptscriptstyle d0}}{3}}\right)\,;\qquad
k_{\scriptstyle eff}= w \left(1- \frac12 R \sqrt{\frac{\Lambda+\kappa \rho_{\scriptscriptstyle d0}}{3}}\right)\,.\een
The first result agrees with the expectations of the cosmological redshift, but the second one is a novel effect.

From the analysis in \cite{EP1} and succinctly described in section \ref{PTA}, we already know that the 
fact that $\sqrt{(\Lambda+\kappa \rho_0)/3}$ is non-zero will produce an enhancement at low values of the angle
subtended between a pulsar and the source, as described.

We postpone a detailed analysis of the application to PTA to a forthcoming publication.


\section{Conclusions and outlook}
In this paper we have reviewed how a non-zero cosmological constant modifies the propagation of a gravitational wave, with
an additional contribution to the one that is usually taken into account ---the redshift in frequency. The effect is entirely 
due to the change of coordinate systems between the reference frame where the wave originates (e.g. the merger of two gigantic
black holes) and the reference frame where waves are measured in PTA, namely cosmological FLRW coordinates.

We have then proceed to extending these results to the case where non-relativistic dust is present, and later to the combined and
more realistic situation where non-relativistic matter and vacuum energy are both present. The results are derived in a 
linearized approximation where only the fist deviations are considered. This approximation is however enough for the case 
of study, as subsequent corrections are seen to be extremely small when the measured value of the matter density and 
cosmological constant are used, combined with the distances involved in the problem.

It came out as a surprise that these deviations were at all measurable, at least in principle, in PTA observations. So far no 
clear positive measurement of GW has come out from the existing collaborations. Perhaps the sort of signal predicted here could help. In
any case the effect is actually reinforced by the presence of non-relativistic matter and the possibility of local measurements of
$\Lambda$ in this way remains.

\section*{Acknowledgments}
We acknowledge the financial support of projects  FPA2013-46570-C2-1-P, FPA2016-76005-C2-1-P and MDM-2014-0369. 
The work of J.A. is partially supported by grants
Fondecyt 1150390 and CONICYT-PIA-ACT14177 (Government of Chile). L.G. acknowledges an FPI grant from MINECO (Spain).

\begin{appendices}
\section{\hspace{-13pt}}
\label{DustDensity}

Let us make a short comment on the equation that relates the density of dust and the variables $t$ and $r$ in \eqref{RelationSolvesODE}. Once the temporal function $C(t)=A\,t^{3/2}$ is fixed, the following change of variables
	\ben\label{Defux}u^3 = r^2\,\kappa\rho\, , \qquad x=A\,\sqrt[3]{\frac{t^2}{r^2}}\een
makes the relation \eqref{RelationSolvesODE} to correspond with the roots of an extended function
	\ben\label{fEq} f(u)=u^3-xu+6\ .\een
The procedure goes as follows: at any particular time $t$, the density profile is determined at each point $r$. The density takes a value such that the function $f$ vanishes, $f(u=\sqrt[3]{r^2\,\kappa\rho})=0$. Or in other words, at a certain point of the spacetime, the relation between the magnitudes $u$ and $x$ is given by the roots of the function $f(u)$.

The new function accomplish $f(-\infty)=-\infty$ and $f(0)=6$, hence at least one of the roots of $f(u)$ has a real and negative value, $u_1<0$. This root of \eqref{fEq} has no physical interpretation because as it has been defined in \eqref{Defux}, $u$ must be positive. Therefore, we know that a positive solution must exist; as
$f(\infty)=\infty$ there must be at least one double root (or  two more roots). To find it (them), let's analyze the function $f(u)$. The critical points are obtained by the roots of the first derivative of $f(u)$
	\ben 
	f'(u_c)=3\,u_c^2-x=0 \qquad\Longrightarrow\qquad u_c=\pm\sqrt{\frac{x}{3}}\,.\een
The second derivative classifies whether the critical points are a maximum, a minimum or a saddle point
	\ben f''(u_c)=6u_c\,.\een
Therefore, being also $x>0$, $u_{c_1}=\sqrt{x/3}$ is a local minimum and  $u_{c_2}=-\sqrt{x/3}$ is a local maximum. We are interested in $u>0$. The next step is to locate the image of the positive critical point $u_{c_1}$ (the minimum), because this value will determine the existence of the wanted remaining root(s)
	\ben f\bigl(\,u_{c_1}=\sqrt{x/3}\,\bigr)=u_{c_1}^3-3\,u_{c_1}^3+6\leq0\,.\een
We are searching for positive real roots of \eqref{fEq}, hence the value of the minimum must be zero or negative (one double root or two simple roots). It is interesting to note that the latter bound can be written as $3^{5/3} \leq x$. Bringing the value of $x$ back, we will see that the previous constraint reflects the presence of a horizon. The mere existence of a positive solution for $f(u)=0$ with $u\propto\rho$ entails the presence of a horizon in such coordinates. With the value of the constant $A=3\sqrt[3]{6}$, the cosmological horizon equation is written as
	\ben\label{HorizonBound}\frac{r}{t}\,\,\leq\,\,\sqrt{\frac{2}{3}}\,.\een

\section{\hspace{-13pt}}
\label{DustEnergyMomentumTensor}
In this appendix we will give the components of the stress-energy tensor explicitly for a dust universe in the new set of coordinates. Dust is a pressureless perfect fluid with an energy-momentum tensor given by \eqref{EMTensorDust}; this definition is coordinate independent. Once the coordinates are fixed one can find the explicit stress-energy tensor by means of Einstein equations \eqref{EinEq}.
The Einstein tensor is straightforward once the metric is chosen; from \eqref{MetricSSSDMUnknown} for instance the computations are particularly simple --it would be equivalent to use the metric \eqref{MetricSSSDM}, at this point $\partial_t\rho_{\scriptscriptstyle dust}$ is already know. 

The product of the two nonnull off-diagonal components of the Einstein tensor lead us to
	\ben\label{EinEqLHS}
	G_t{}^r\,G_r{}^t=-\frac{(\kappa\rho_{ \scriptscriptstyle dust})^3\,r^2} {3\left(1-\frac{\kappa\rho_{ \scriptscriptstyle dust}}{3}\,r^2\right)^2}\,.\een
Notice that $G_t{}^r\neq G_r{}^t$. The r.h.s. of Einstein equations \eqref{EinEq} equals the previous relation to
	\ben\label{EinEqRHS}
	\kappa^2\,T_t{}^r\,T_r{}^t=(\kappa\rho)^2\,U_tU^t\,U_rU^r\,.\een
We aim at solving the 4--velocity and we have another equation to do so: the normalization condition over the 4--velocity that yields
	\ben\label{Norm4Vel}
	g_{\mu\nu}\,U^\mu U^\nu=U_tU^t+U_rU^r=1\,.\een
The system of two equations: \eqref{EinEqLHS}=\eqref{EinEqRHS} and \eqref{Norm4Vel} has the following solution
	\begin{align}\label{VelocityTemp}
	&U_tU^t=\frac{1}{1-\frac{\kappa\rho_{ \scriptscriptstyle dust}}{3}\,r^2}\,;\\&U_rU^r=-\frac{\kappa\rho_{ \scriptscriptstyle dust}}{3}\,\frac{r^2}{1-\frac{\kappa\rho_{ \scriptscriptstyle dust}}{3}\,r^2}\,.\end{align}
We have chosen the solution noticing that $U_tU^t>0$ and $U_rU^r<0$. The next step is to isolate each component of the velocity; using the prescription $U_\mu^2=U_\mu U^\mu\,g_{\mu\nu}$, for $\mu=t,r$ we obtain
	\begin{align}
	&U_t=\frac{\abs{\partial_t\rho_{ \scriptscriptstyle dust}}}{(3\,\kappa\rho_{ \scriptscriptstyle dust}^3)^{1/2}}\,;\\ &U_r=\sqrt{\frac{\kappa\rho_{ \scriptscriptstyle dust}}{3}}\,\frac{r}{1-\frac{\kappa\rho_{ \scriptscriptstyle dust}}{3}\,r^2}\,.
	\end{align}

A relevant comment is that if the temporal Einstein equation
	\ben
	G_t{}^t=\kappa\,\frac{r\,\partial_r\rho_{ \scriptscriptstyle dust}+3\,\rho}{3}=\kappa\,\rho\,U_tU^t\een
is combined with the corresponding velocity in Eq. (\ref{VelocityTemp}), and then one solves for the radial derivative 
one gets (\ref{DiagonalMetricCondition}).

Finally
	\begin{equation}
	\begin{aligned}
	&T_t{}^t=\frac{3\,\rho_{\scriptscriptstyle dust}}{3-\kappa\rho_{ \scriptscriptstyle dust}\,r^2}&&&&\qquad T_t{}^r=-\frac{\abs{\partial_t\rho_{ \scriptscriptstyle dust}}}{3}\,r
	\\[1ex]
	&T_r{}^t=\frac{9\,\kappa\rho_{ \scriptscriptstyle dust}^3}{\abs{\partial_t\rho_{ \scriptscriptstyle dust}}\left(3-\kappa\rho_{ \scriptscriptstyle dust}\,r^2\right)^2}
	\,r&&&&\qquad T_r{}^r=-\frac{\kappa\rho_{ \scriptscriptstyle dust}^2}{3-\kappa\rho_{ \scriptscriptstyle dust}\,r^2}\,r^2
	\end{aligned}\end{equation}

\end{appendices}

\end{document}